# A circuit for precise random frequency synthesis via a frequency locked loop

Mario Stipčević[1*]

[1]Center of excellence for advanced materials and sensors, Ruđer Bošković Institute, Bijenička 54, HR-10000 Zagreb, Croatia

**Frequency synthesis (FS) is a technique vital for all kinds of radio frequency (RF) communications, such as: mobile phones, Bluetooth, Wi-Fi, radio, TV and satellite, and in other equipment requiring periodic signals of stable and programmable frequency. In this work, we present a generalization of the FS technique to random, non-periodic signals, whose main use is in the new area of stochastic neuromorphic computing and, information security and instrumentation. Since conventional FS circuits cannot work with random signals, we introduce a novel random frequency ratio detector, that works both with random and periodic signals.**

Frequency synthesizer is an electronic circuit capable of generating a periodic signal of stable frequency in a certain frequency interval. The set of available frequencies is discrete, with frequencies separated by an arbitrarily small, constant spacing. Frequency synthesizers are crucial for operation of many widely used devices, such as: computers, two-way radio communication devices (e.g. cell phones, Bluetooth, Wi-Fi, walkie-talkies), receivers and transmitters (for radio, TV, and radio controlled devices), global positioning system (GPS), satellite communications, laboratory equipment and many other applications. Frequency synthesis (FS) of periodic signals can be accomplished by well-known methods of phase-locked loop (PLL) [3] or direct digital synthesis (DDS) [4]. These methods rely on one or more basic techniques, notably: frequency division, signal multiplication and mixing, phase detection, phase frequency detection (PFD), low-pass filtering, voltage controlled oscillators (VCO) and digital-to-analog conversion.

In particular, PLLs are widely employed because of their good frequency stability and low phase noise and because, on top of FS, they can be used to demodulate a signal, recover a signal from a noisy communication channel, generate a stable frequency at multiples of an input frequency, or distribute precisely timed clock pulses in digital logic circuits such as microprocessors. Since a single integrated circuit can provide a complete PLL building block, the technique is widely used in modern electronic devices, with output frequencies from a fraction of a hertz up to many gigahertz. Signals obtained from FS are typically sinusoidal or square-wave with duty cycle of (roughly) ½.

In this work, however, we are concerned with a different type of electrical signal, namely random pulse train (RPT). By definition, a RPT is a sequence of square logic pulses of a constant voltage amplitude and typically, but not necessarily, of a constant duration $\tau_p$, as shown in Fig. 3, wherein time intervals $t_i$ between rising edges of subsequent pulses (also known as "waiting times") are random numbers.

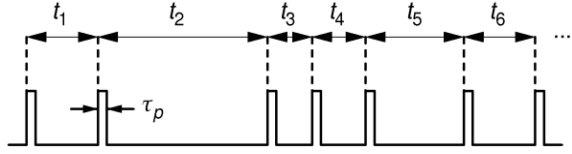

**Figure 3.** A random pulse train (RPT) is a sequence logic pulses usually of constant duration $\tau_p$. Shown is a voltage variation of a RPT as a function of time. For a Poissonian RPT, pulse waiting times $t_i$ ($i$=1,2,3,...) are random variates drawn from the exponential probability density function.

We note that, by convention, the important information in a RPT is contained in rising edges: falling ones are of no consequence, except technical. In this work, unless specifically mentioned, we deal with Poissonian RPT, wherein time intervals $t_i$ follow the exponential probability density function (p.d.f.):

$$P(t) = \frac{1}{\tau} e^{-t/\tau}. \tag{1}$$

Thus, a RPT is characterized by a single parameter: the average waiting time, $\tau$, whose inverse is henceforth referred to as a "random frequency" or just "frequency" and denoted by:

$$f_R = \frac{1}{\tau}. \tag{2}$$

It is best to picture a RPT as being a result of an underlying stationary and memoryless physical process wherein each event coincides with a rising edge of a logic pulse. Quite generally, for empirical signals, random or periodic, we define frequency as the inverse of the average waiting time:

$$f_R = \frac{1}{\lim_{N \to \infty} \frac{1}{N} \sum_{i=1}^{N} t_i}. \tag{3}$$

For a RPT based on a Poissonian random process, this definition coincides with Eq, 2.

**Phase and frequency locked loops**

This work is about a means for generation of a RPT of a precise frequency. Previous art [8] relates to systems in which the output frequency is controllable via an input control voltage. This is conceptually similar to a well-known voltage-controlled oscillator (VCO) used in periodic frequency generation, except that here output is a RPT, instead of a periodic signal. The main problem with such a system is that it does not possess a compensating mechanism and therefore its parameters, and thus the set random frequency, can easily adrift from a desired value with time, temperature or aging. In order to get a stable frequency, we propose to use a control loop control similar to a PLL.

A PLL is a negative-feedback control system that generates an output signal whose phase is synchronized to the phase of a very stable local oscillator (LO) (usually a quartz-controlled oscillator). There are several types of a PLL; the simplest is an electronic circuit consisting of a variable frequency oscillator (VCO), a programmable divider-by-n divider and a phase detector (PD) followed by a DC level extractor, in a negative-feedback loop, as shown in Fig 7. A principle is simple: a periodic signal from VCO divided by an integer number and signal from LO are fed into the phase detector which generates an "error voltage" proportional to the phase difference of its input signals.

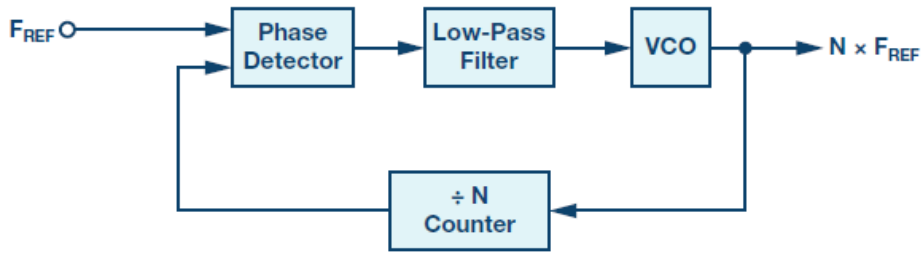

**Figure 7.** A basic PLL circuit.

The error voltage is fed back to the VCO such that its frequency is adjusted until the phases (and frequencies) match. A simple XOR gate, shown in Fig. 8a suffices to detect phase difference of two square-wave signals of the same duty cycles and is thus frequently used in PLLs. However, the problem with XOR PD is that is can generate local minima when one input frequency is an integer multiple of the other and therefore a "lock" can happen at an undesired frequency, and for a correct operation it is required that input signals have the same duty cycle. To improve on both accounts, an improved circuit shown in Fig. 8b has been proposed in seventies and is used ever since [9]. In this PD, error voltage is an additive combination of differences in both frequency and phase, of the input signals. Therefore its name "phase frequency detector" (PFD). Because of the edge-triggered action, duty cycles of the input signals are irrelevant.

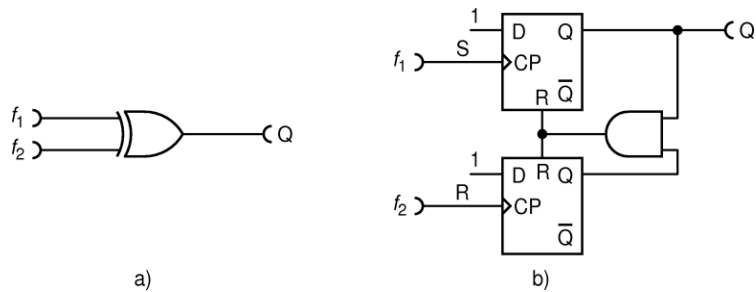

**Figure 8.** A simple XOR-based phase detector (a); a classic phase frequency detector (b).

...... In conclusion, the average duty cycle at the output of XOR and PFD circuits has no sensitivity to frequency of the RPT whatsoever. It is easy to see that the same conclusion holds for a RPT/2D signal as well which has an average duty cycle of ½. The reason for that is that these circuits have been constructed primarily to measure phase difference between a pair of input signals, having some sensitivity to a frequency difference as a collateral bonus, while relative phase of two Poissonian random signals contains no information whatsoever.

Therefore, we need to construct a different frequency detector for purpose of random frequency synthesis via a FLL. We propose a well-known clock-less RS flip-flop but with a twist that it has the edge action implemented on both inputs, as shown in Fig. 9.

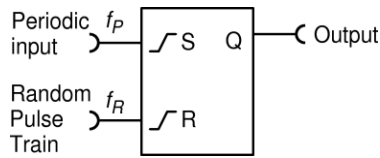

**Figure 9.** A proposed frequency ratio detector (FRD).

This circuit acts asymmetrically: if the two inputs are switched, its response function mirrors horizontally around the point $f_R = f_P$.

Unfortunately, the clock-less edge-triggered RS flip flop is not available as a standard electronic component. We have to construct it from other logic circuits. Notably, the D-type flip-flop with an edge-triggered clock input presents an industry standard and is widely available, for example in FGPAs. The new circuit, that we name "frequency ratio detector" (FRD), made of two D-type flip-flops, equivalent to the modified RS flip-flop, is shown in Fig. 10.

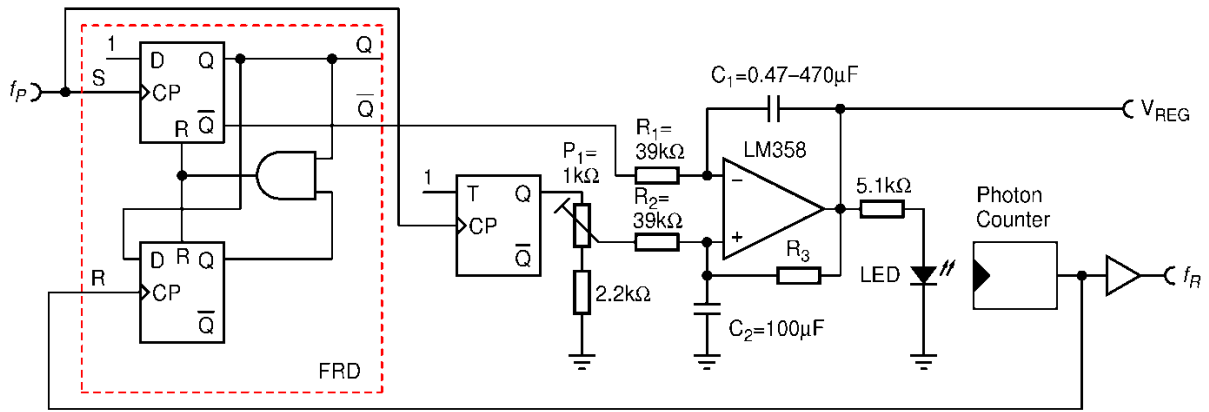

**Figure 10.** Frequency ratio detector (FRD) in a frequency-locked loop capable of generating RPT in a range from 200 cps to 20 MCps. Output frequency $f_R$ is locked (in 1:1 ratio) onto the frequency of the periodic local oscillator of frequency $f_p$.

Let us now imagine that a periodic signal, of frequency $f_P$, is sent to the input S, and that a RPT of an average frequency $f_R$ is sent to the input R. A typical waveform present at the output Q of the FRD is shown as a solid (blue) line in Fig. 11. Dashed line shows the periodic signal present at the input S, having the period $T = 1/f_P$. The output Q goes HIGH by each rising edge of the periodic input, if not already HIGH in which case the input S is ignored. The output Q goes low by each rising edge of the (random) signal connected to the input R, if not already LOW in which case the input R is ignored.

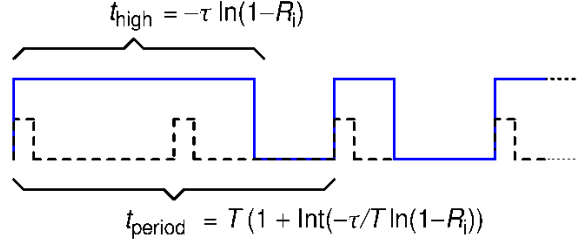

**Figure 11.** An example of the waveform at the output Q of a FRD is shown by the solid (blue) line. Periodic signal at the input R is shown by a dashed line. Random events correspond to falling edges of the solid line.

It is important to note that whenever one starts to measure the time (for example at a time of rising edge of the periodic signal that brings output of the RFFL at HIGH state), the next random event will happen after a time delay which is a random variate drawn from exponential p.d.f. with the mean value $\tau = 1/f_R$, and therefore durations of the HIGH state will be such variates. Let us now define a "period" as a time lapse between two subsequent rising edges at the output Q. Then, the total duration of the HIGH states of the output Q during $N$ periods is a sum over all periods:

$$T_{\text{high}} = \sum_{i=1}^{N} -\tau \cdot \ln(1 - R_i) \tag{4}$$

where $T = 1/f_P$, $\tau = 1/f_R$, and $R_i$ is $i$-th random number drawn from the uniform probability distribution. The total duration of $N$ periods, each period consisting of a HIGH state followed by a LOW state, is a sum of all periods:

$$T_{\text{tot}} = T \sum_{i=1}^{N} (1 + \text{Int}\left(-\frac{\tau}{T} \cdot \ln(1 - R_i)\right). \tag{5}$$

Now, the average duty cycle is defined as:

$$\langle D \rangle = \lim_{N \to \infty} \frac{T_{\text{high}}}{T_{\text{tot}}} = \frac{\lim_{N \to \infty} \frac{1}{N} \sum_{i=1}^{N} -\tau \cdot \ln(1 - R_i)}{T \left(1 + \lim_{N \to \infty} \frac{1}{N} \sum_{i=1}^{N} \text{Int}\left(-\frac{\tau}{T} \cdot \ln(1 - R_i)\right)\right)}. \tag{6}$$

The numerator term is just the average period, $\tau$, so the Eq. 6 can be written as:

$$\langle D \rangle = \frac{\tau/T}{1 + \lim_{N \to \infty} \frac{1}{N} \sum_{i=1}^{N} \text{Int}\left(-\frac{\tau}{T} \cdot \ln(1 - R_i)\right)}. \tag{7}$$

The sum in the denominator can be rearranged as a sum of terms in which real positive numbers $-\frac{\tau}{T} \cdot \ln(1 - R_i)$ lie in an interval between two consecutive integer numbers:

$$\lim_{N\to\infty}\frac{1}{N}\sum_{i=1}^{N}\text{Int}\left(-\frac{\tau}{T}\cdot\ln(1-R_i)\right)= \tag{8}$$

$$=0\int_{0}^{T/\tau}e^{-z}dz+1\int_{0}^{2T/\tau}e^{-z}dz+2\int_{0}^{3T/\tau}e^{-z}dz+\cdots \tag{9}$$

$$=\sum_{k=1}^{\infty}k\int_{k}^{(k+1)T/\tau}e^{-z}dz=\sum_{k=1}^{\infty}ke^{-kT/\tau}\left(1-e^{-kT/\tau}\right) \tag{10}$$

$$=\left(e^{-T/\tau}-e^{-2T/\tau}\right)+2\left(e^{-2T/\tau}-e^{-3T/\tau}\right)+3\left(e^{-3T/\tau}-e^{-24}\right)+\cdots \tag{11}$$

$$=\sum_{k=1}^{\infty}e^{-kT/\tau}=\frac{e^{-T/\tau}}{1-e^{-T/\tau}}. \tag{12}$$

Finally, inserting Eq. 12 into Eq. 7 we get:

$$\langle D\rangle=\frac{\tau}{T}\left(1-e^{-T/\tau}\right). \tag{13}$$

Since the average duty cycle, $\langle D\rangle$, is a function only of the frequency ratio of the two input signals, namely $T/\tau=f_R/f_P$, the circuit shown in Fig. 10 can be used as a frequency detector. Its transfer function, given in Eq. 11, along with the actual measured points, is shown in Fig 12.

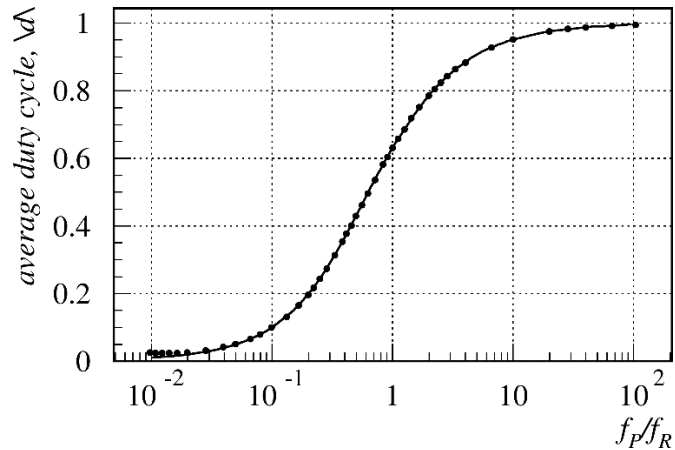

**Figure 12.**

In theory, this circuit is accurate in the sense that..... However, in practice, the average voltage at V- is not perfectly proportional to the average frequency of FD Output because of the finite rise and fall times of the digital signal. Shorter pulses will suffer greater errors due to the edge effects in proportion to their length. The net effect of this is that as $f_{\text{in}}$ rises, the $f_{\text{out}}$ lags behind. In our particular circuit the error reaches -0.88% at $f_{\text{in}}=20$ MHz. This error is linear in the average pulse length at the FRD Output and therefore can be corrected by injecting a small current to positive input of the operational amplifier

LM358 proportional to the input frequency $f_{in}$. To that end, assuming approximately linear relationship between the voltage $V_0$ and light intensity of the LED, we simply use a feedback resistor $R_3 = 2.2$ M$\Omega$, shown on Fig. 10 whose optimal value is determined by experiment such as to minimize the non-linearity.

None of the PDs or FPDs known so far work if one or both inputs is a RPT. This is interesting because noone has constructed them sepecifically to be insensitive to random signals, but they are. On the other hand, circuit in Fig. 10 works well when one signal is periodic and the other random, when both are random and even when both are periodic. In fact, when fed by two periodic signals it also features a level of phase sensitivity without a gap, and thus presents an improvement and generalization of FPDs known so far.

**Frequency precision**

Resulting linearity of the frequency $f_R$ of the output RPT, versus frequency $f_P$ of the periodic control input, is shown in Fig. 13. We see that the relative error is kept well within 1% over 5 orders of magnitude.

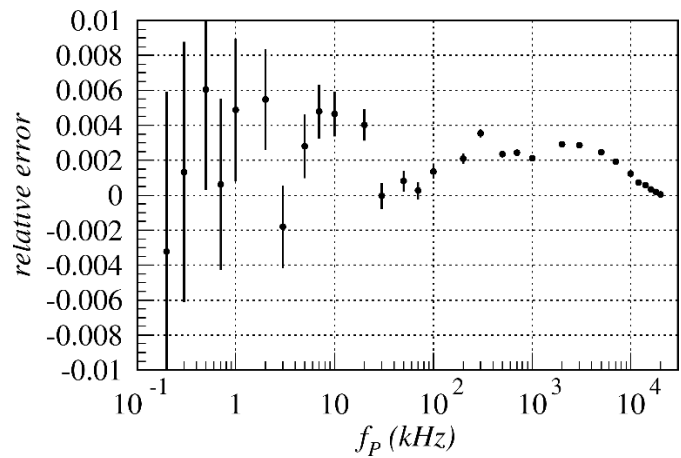

**Figure 13.**

**Frequency jump response**

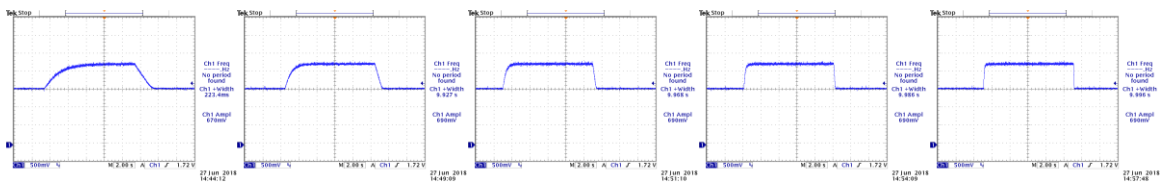

**Figure 14.** C1=110 uF, 47 uF, 22.5 uF, 10.5 uF, 5.1 uF , modulation frequency = 0.05 Hz, pulse duration = 10s, scope resolution 2 s/div. (Nejasno je značenje 100k i 1M.)

**Randomness quality and autocorrelation**

Distributions / fine. Autocorrelation is sensitive o small randomness imperfections:

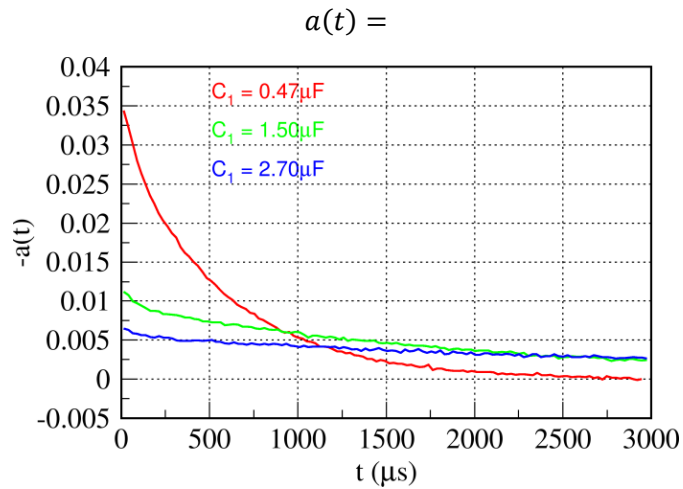

$$a(t) =$$

**Figure 14.**

Autocorrelation function a(k) depends on mean random period tau and gain and time constant R1C1 of the feedback loop. ne ovisi o tau (1/frekvencija?) ako se prika\e kao funkcija a(t) gdje t=k*tau.

**Modulation and demodulation**

Spomenuti i brzu verziju s DRSF koja kao ulaz treba jednu ili dvije stabilne točne frekvencije, Ref. [7].

A system for generating random frequencies in a desired range with an arbitrarily small increment, named digital random frequency synthesis (DRFS), has been described in our earlier work [7]. However, this requires two stable sources of precisely known random frequencies - a problem that was not solved at the time.